\documentclass[prb,twocolumn,preprintnumbers,superscriptaddress]{revtex4-2}

\usepackage{amsmath}
\usepackage{xcolor}
\usepackage{graphicx}
\usepackage{dcolumn}
\usepackage{bm}
\usepackage{hyperref}

\begin{document}


\title{Exchange interaction for Mn acceptor in GaAs: \\ revealing its strong deformation dependence}

\author{I.~V. Krainov}
\email{igor.kraynov@mail.ru} 
\affiliation{Ioffe Institute, 194021 St.~Petersburg, Russia} 
\author{K.~A. Baryshnikov}
\email{barysh.1989@gmail.com}
\affiliation{Ioffe Institute, 194021 St.~Petersburg, Russia} 
\author{A.~A. Karpova}
\affiliation{Ioffe Institute, 194021 St.~Petersburg, Russia} 
\affiliation{Saint-Petersburg Electrotechnical University, 197022 St.~Petersburg, Russia}
\author{N.~S. Averkiev}
\affiliation{Ioffe Institute, 194021 St.~Petersburg, Russia} 

\date{\today}

\begin{abstract}
In this paper we calculate exchange interaction constant between manganese ion inner electronic
$d$-shell and GaAs valence band bounded hole using their microscopic multiparticle wave functions.
We reveal its parametric dependence on crystal lattice deformations and find out that it could be
about and even more than dozens percent when the strain tensor reaches values of $10^{-3} \div 10^{-2}$. This fact is in accordance with the previous hypothesis of deformation dependence of Mn acceptors in GaAs fine energy structure obtained from Raman spectroscopy, and we show that this dependence has the same magnitude. Also, we resolve here the problem of a substantial high temperature
mismatch between well-developed theory and experimental data for the static magnetic susceptibility
of Mn ions in GaAs. We show by numerical estimates and calculations that quite a strong parametric
dependence of the exchange coupling value on GaAs lattice expansion determines the high
temperature (above $50$~K) magnetic susceptibility reduction as well.

\end{abstract}

\maketitle


\section{Introduction}

Modern material science is focused on functional materials combining different properties with maximal functionality. One of these important kinds of such materials is magnetic semiconductors mixing electrical, optical and magnetic properties. Different ways to control these properties merge into the important directions of research, including the production of new compounds \cite{awschalom2007challenges,ortiz2019new,otrokov2019unique,gibertini2019magnetic,wang2020prospects,jena2019new}, nanostructure design \cite{awschalom2007challenges,tsymbal2019spintronics,gibson2015three,mak2022semiconductor,need2020magnetic,furdyna2012exchange} and investigation of the effects of external forces application \cite{mogi2022experimental,mccreary2020quasi,mak2019probing,krainov2021spin,averkiev2018mnga}. One of the most well-known functional materials is GaMnAs \cite{lee2009ferromagnetic,jungwirth2006theory,yuan2017interplay,liu2006ferromagnetic}. In this material manganese ion with its inner magnetic $3d$-shell containing 5 electrons brings magnetism to GaAs semiconductor host. This is due to the exchange interactions between manganese inner $d$-electrons with GaAs holes.
Also, the manganese impurity acts as an acceptor increasing hole concentration in GaAs semiconductor crystal. For an isolated impurity the exchange interaction between Mn half-filled $d$-shell with the total spin of electrons $5/2$ and localized hole in the $\Gamma_8$ symmetry state acting like a $3/2$ spin results in initially $24$-hold degenerate state into into $4$ sublevels with total angular momentum $F = 1,2,3,4$ with $F = 1$ being the ground state \cite{averkiev2018mnga}. Here we stress our attention on this exchange interaction between isolated manganese ion and a hole bounded on it.  

Exchange interaction constant $A$ for Mn acceptor consist of two parts. The first part includes the exchange between Mn $d$-shell orbital electrons and Bloch orbital of the $\Gamma_8$ hole, and the second part includes value of hole envelope at impurity site, i.e., the probability to couple with the half-filled $d$-shell as a whole. In all previous works \cite{yakunin2007warping,MONAKHOV2008416,nestoklon2015fine}, in which such interaction was discussed, only the second part (value of hole envelope at impurity site) was assumed to change in different conditions, while the first part (exchange between Bloch functions) was assumed to be unperturbed, and its value was postulated \cite{averkiev2018mnga}. The deformation influence on the envelope part of exchange constant has been investigated in \cite{nestoklon2015fine}, but it has been found that it changes by less than one percent at pressures on the limit of GaAs hardness. The purpose of this work is to calculate exchange interaction value between Bloch functions of Mn $d$-shell and $\Gamma_8$ hole bounded from the GaAs valence band, and to treat its dependence on deformation. We demonstrate that this part of exchange interaction is sensitive to the presence of crystal strains. 

Previously, the assumption of a strong dependence of exchange constant $A$ on the crystal deformation played a crucial role in the study of the fine structure of an isolated Mn acceptor in GaAs. The latter was investigated using Raman spin-flip scattering and its dependence on magnetic fields and external deformations at helium  temperatures \cite{Krainov2016}. The theoretical fit of intra and inter transitions between Mn-hole levels based on a standard model of the Mn acceptor eigenstates was also carried out in \cite{Krainov2016}, but to make a satisfactory agreement between all experimental curves and theoretical calculations the deformation dependence of exchange interaction constant was phenomenologically proposed and its value was estimated from comparison with the experiment. The exchange interaction value changes by $20$~\% for the pressure $5$~kbar, which is about half of GaAs critical value of hardness, and hence this change is much larger than previously mentioned nearly one percent dependence on hole envelope wave function change.  


Independently, there are drastically different measurements of static magnetic susceptibility behavior in a wide temperature range in GaAs samples with low concentration of Mn ions. The first experiments were made by Andrianov's group \cite{andrianov1983characteristics}, but their work contains an irrelevant theoretical model of the Mn center, which mismatches with a bunch of low temperature properties of the center. Other measurements were carried out by Frey's group and reported in \cite{frey1988paramagnetism}, where the relevant theoretical model was applied, which, however, has some discrepancies with the data at the very high temperature edge. The state of art of these studies is that the theoretical fit based on that true and now standard Mn-hole interaction model of experimental data is in a good agreement with the low temperature region below $50$~K. But for the high temperature region, there is a reduction of magnetic susceptibility compared with the theoretical prediction, which is still puzzling nowadays. A recent paper \cite{averkiev2018mnga} containing a deep review of different experimental and theoretical facts about Mn center in GaAs proposed a hypothesis that variance mentioned above could be explained by the Jahn–Teller effect (JTE). 

In this paper we also test this hypothesis (see the Supplementary material). It is known from many other experimental facts \cite{averkiev2018mnga} that the Mn ground $F = 1$ state is unaffected by the static Jahn–Teller distortion, so only dynamical JTE should be tested \cite{JTBersuker}. Moreover, one can show that at high temperatures, there is only one way for dynamical JTE to occur, which is reduced to the Jahn-Teller interaction of hole in $\Gamma_8$ state with local lattice distortions.
As we show (see the Supplementary material, part 2), the high temperature dependence of magnetic susceptibility is negligibly dependent on the Jahn–Teller effect and ceases quite rapidly as temperature increases that can not explain observed reduction of magnetic susceptibility discussed above.  
Also we test a hypothesis of the crystal field influence, but it also can not explain magnetic susceptibility reduction at high temperatures (see the Supplementary material, part 1). 
But here we show that if we link the phenomenological dependence of exchange interaction value on external deformation from \cite{Krainov2016} with the thermal expansion coefficient of the crystal, the problem of high-temperature magnetic susceptibility reduction can be elegantly resolved.

In this paper we will calculate Mn-hole exchange-interaction value part associated with the Bloch wave functions overlapping. Then we provide an estimate for this strain dependence. Note that the trace of strain tensor for the pressure about $5$~kbar is in the range of $10^{-3}$ -- $10^{-2}$, and it is quite surprising how it can lead to a strong dependence of the exchange constant $\sim 20\%$. We elaborate and explain a simple mechanism that could explain this fact. 
Further, we show that such purely theoretical estimates result in the similar variation for Mn-hole exchange constant on stress as assumed in \cite{Krainov2016}, which has the same order of value. 
Finally, we show by direct calculations that the obtained dependence of $A$ on the crystal strains $\varepsilon$, which theoretically fits Mn fine energy structure \cite{Krainov2016}, leads to a better agreement between high-temperature magnetic-susceptibility calculation results and experimental data.
We also believe that the developed model could be applied to another magnetic impurities and hosts with appropriate modifications in symmetry analysis.

\section{Theory}
\subsection{Exchange Hamiltonian and representation of total angular momenta F = 1, 2, 3, and 4.}

Eigenstates of Mn acceptor are composed from the sixfold degenerate state of Mn ion $d$-shell electrons in the ground state with total spin $S=5/2$ and the fourfold degenerate state of a localized hole having the $\Gamma_8$ symmetry, which corresponds to the total angular momentum $J=3/2$. 
Further, to simplify all conclusions, we will work in the hole basis of the $d$-shell, which has the same properties as the electronic one, because the shell is half-filled, and one-particle states simply have opposite spins. 
These eigenstates are split by exchange interaction between the half-filled $d$-shell and the localized hole resulting in the total angular momentum states $F=1,2,3,4$ with corresponding degeneracy equal to $2F+1$. 

So, if we assume that the exchange interaction between the ion's $d$-shell and the hole is described by the only one constant $A$, i.e., if we set the corresponding Hamiltonian as
\begin{gather}
  \label{H_exchange}
    \displaystyle \hat{H}_{\rm ex} = A ( \hat{\bm S} \cdot \hat{\bm J} ) = \frac{A}{2}\left( \hat{F}^2 - \hat{S}^2 - \hat{J}^2 \right), \\ \nonumber
    \hat{F} = \hat{S} + \hat{J},  
\end{gather}
then one can easily find out all its energy eigenvalues, which are $A(F(F+1) - S(S+1) - J(J+1))/2$. 
All other possible terms proportional to $( \hat{\bm S} \cdot \hat{\bm J} )^2$ and $( \hat{\bm S} \cdot \hat{\bm J} )^3$ are connected with the second-order and higher-order perturbation terms of Coloumb interaction causing change of spin projections of $d$-shell electrons. 
We will neglect such terms because the energy of spin-spin interaction between the $d$-shell electrons is assumed to be the largest among all other energies. This assumption allows us to consider all processes as if no changes in spin states of the inner shell electrons occur.
Note also that there are no spin-orbit splittings in the $d$-shell, which is confirmed by the Raman scattering data of Mn$^{0}$ centers in GaAs, which has g-factor strictly equal to $2$ \cite{Krainov2016, Cardona}.

To calculate the eigenenergies of $\hat{H}_{\rm ex}$, it is sufficient to use the subset from the whole basis of acceptor states, because of the spherical symmetry of the Hamiltonian. Let us consider such a subset consisting of only $4$ wave functions $|F,F_z = 0 \rangle$ (where $F=1,2,3,4$), and taking it from \cite{averkiev2018mnga} (note that the prefactor coefficient in $|2,0\rangle$ function is changed to normalize correctly the wave function) one can write it down as
\begin{gather}
\label{Fz0_states} \nonumber
    \displaystyle |1,0\rangle = \frac{1}{\sqrt{5}} \biggl\{ \Psi^S_{3/2} \Psi^J_{-3/2} - \Psi^S_{-3/2} \Psi^J_{3/2} - \\ 
    - \sqrt{\frac{3}{2}} \Psi^S_{1/2} \Psi^J_{-1/2} 
    + \sqrt{\frac{3}{2}} \Psi^S_{-1/2} \Psi^J_{1/2} \biggl\}, \\ \nonumber
    \displaystyle |2,0\rangle = \sqrt{\frac{3}{7}} \biggl\{ \Psi^S_{3/2} \Psi^J_{-3/2} + \Psi^S_{-3/2} \Psi^J_{3/2} - \\ 
    - \sqrt{\frac{1}{6}} \Psi^S_{1/2} \Psi^J_{-1/2} 
    - \sqrt{\frac{1}{6}} \Psi^S_{-1/2} \Psi^J_{1/2} \biggl\}, \\ \nonumber
    \displaystyle |3,0\rangle = \frac{1}{\sqrt{5}} \biggl\{ \sqrt{\frac{3}{2}} \Psi^S_{3/2} \Psi^J_{-3/2} - \sqrt{\frac{3}{2}} \Psi^S_{-3/2} \Psi^J_{3/2} + \\ 
    + \Psi^S_{1/2} \Psi^J_{-1/2} 
    -  \Psi^S_{-1/2} \Psi^J_{1/2} \biggl\},  \\ \nonumber
    \displaystyle |4,0\rangle = \sqrt{\frac{3}{7}} \biggl\{ \sqrt{\frac{1}{6}} \Psi^S_{3/2} \Psi^J_{-3/2} + \sqrt{\frac{1}{6}} \Psi^S_{-3/2} \Psi^J_{3/2} + \\
    + \Psi^S_{1/2} \Psi^J_{-1/2} 
    + \Psi^S_{-1/2} \Psi^J_{1/2} \biggl\}.
\end{gather}
Then one can calculate all energy differences between $\hat{H}_{\rm exch}$ eigenstates as 
\begin{gather}
\nonumber
E_{F + 1} - E_{F} = \\
\nonumber
= \langle F+ 1,0 |\hat{H}_{\rm exch}| F+1,0 \rangle - \langle F,0 |\hat{H}_{\rm exch}| F,0 \rangle = \\
= 2A,~3A,~4A.
\label{energy_diff}
\end{gather}
This result could be obtained by taking subset of $4$ wave functions, which contain only zero projections of the total momentum on $z$ axis: $\left\{ \Psi^S_{3/2} \Psi^J_{-3/2} \right.$; $\Psi^S_{-3/2} \Psi^J_{3/2}$; $\Psi^S_{1/2} \Psi^J_{-1/2}$; $\left. \Psi^S_{-1/2} \Psi^J_{1/2} \right\}$, generating $|F,0\rangle$ states. By calculating the exchange Hamiltonian using these wave functions as bra and ket functions, one can obtain a $4\times 4$ matrix, which eigenvalues give us the same energy differences as in Eq.~(\ref{energy_diff}). 

So, the main idea for microscopic calculation of $A$ via exchange integrals is to consider the first-order correction to the energies of $d$-states and of the hole state due to the Coulomb interaction calculated using only these $4$ wave functions with appropriate symmetrization and antisymmetrization of all multiparticle orbitals and spin states.


\subsection{Microscopic calculation of exchange integrals.}
Wave functions of bounded $\Gamma_8$ hole corresponding to the total moment $J = 3/2$ include envelope and Bloch parts. Within the framework of the effective mass method for shallow acceptors in cubic semiconductors in the spherical approximation, the wave function of this hole is the sum of the products of the Bloch amplitudes $X_{\mu}$ and the smooth envelopes $R_0(r)$ and $R_2(r)$
\begin{eqnarray}
    \label{envelope_functions}
    \displaystyle \Psi^J_{3/2} = R_0(r) Y_{00} X_{3/2} + \frac{1}{\sqrt{5}} R_2(r) Y_{20} X_{3/2} - \nonumber \\ 
    \displaystyle - \sqrt{\frac{2}{5}} R_2(r) Y_{21} X_{1/2} + \sqrt{\frac{2}{5}} R_2(r) Y_{22} X_{-1/2}, \\
    \displaystyle \Psi^J_{1/2} = R_0(r) Y_{00} X_{1/2} - \frac{1}{\sqrt{5}} R_2(r) Y_{20} X_{1/2} + \nonumber \\ 
    \displaystyle + \sqrt{\frac{2}{5}} R_2(r) Y_{2,-1} X_{3/2} + \sqrt{\frac{2}{5}} R_2(r) Y_{22} X_{-3/2},  \\
    \displaystyle \Psi^J_{-1/2} = R_0(r) Y_{00} X_{-1/2} - \frac{1}{\sqrt{5}} R_2(r) Y_{20} X_{-1/2} + \nonumber \\ 
    \displaystyle + \sqrt{\frac{2}{5}} R_2(r) Y_{2,1} X_{-3/2} + \sqrt{\frac{2}{5}} R_2(r) Y_{2,-2} X_{3/2},  \\
    \displaystyle \Psi^J_{-3/2} = R_0(r) Y_{00} X_{-3/2} + \frac{1}{\sqrt{5}} R_2(r) Y_{20} X_{-3/2} - \nonumber \\ 
    \displaystyle - \sqrt{\frac{2}{5}} R_2(r) Y_{2,-1} X_{-1/2} + \sqrt{\frac{2}{5}} R_2(r) Y_{2,-2} X_{1/2}, ~
\end{eqnarray}
where $Y_{lm}$ are the spherical functions corresponding to the orbital moment $l$ and its projection $m$. The exchange interaction integral will involve these functions and the $d$-shell wave functions, which are located in one elementary cell at the impurity site. We can neglect the effect of $R_2(r)$ functions because they tend to zero limit at the magnetic impurity site, while $R_0(r)$ functions take nonzero values (see calculations results in \cite{averkiev2018mnga}). Thus, in App.~\ref{App1} we calculate all exchange integrals using only Bloch parts of $\Gamma_8$ hole wave functions, setting $\Psi^J_{\mu} \approx f(0) X_{\mu}$, where $f(0) = R_0(0)/\sqrt{4\pi}$.

Basing on spin configurations of wave functions $\psi_i$ ($i=1,2,3,4$)
\begin{gather}
\label{Basis_functions}
    \left\{ \Psi^S_{3/2} \Psi^J_{-3/2};~\! \Psi^S_{-3/2} \Psi^J_{3/2};~\! \Psi^S_{1/2} \Psi^J_{-1/2};~\! \Psi^S_{-1/2} \Psi^J_{1/2} \right\} \qquad
\end{gather}
one can show that the Hamiltonian of Coulomb interaction has a following $4\times 4$ matrix form in this basis
\begin{equation}
\label{Coloumb_Hamiltonian_4x4}
    \displaystyle \hat{H}_{\rm C} = 
    \begin{pmatrix}
        \mathcal{X} & \mathcal{Z} & 0 & 0 \\
        \mathcal{Z} & \mathcal{Y} & \mathcal{V} & 0 \\
        0 & \mathcal{V} & \mathcal{Y} & \mathcal{Z} \\
        0 & 0 & \mathcal{Z} & \mathcal{X} 
    \end{pmatrix} .
\end{equation}
The details of calculation one can see in App.~\ref{App1}, where the true microscopical multiparticle structure of all wave functions is taken into account, here we represent the results
\begin{gather}
\nonumber
    \displaystyle \mathcal{X} = \langle \Psi^S_{3/2} \Psi^J_{-3/2} | U ({\bm r_1} - {\bm r_2}) | \Psi^S_{3/2} \Psi^J_{-3/2} \rangle = \\ \label{X}
    = W + \ae |f(0)|^2, \\
\nonumber
    \displaystyle \mathcal{Y} = \langle \Psi^S_{1/2} \Psi^J_{-1/2} | U ({\bm r_1} - {\bm r_2}) | \Psi^S_{1/2} \Psi^J_{-1/2} \rangle = \\ \label{Y}
    = W + \frac{7}{3} \ae |f(0)|^2, \\
\nonumber
    \displaystyle \mathcal{Z} = \langle \Psi^S_{3/2} \Psi^J_{-3/2} | U ({\bm r_1} - {\bm r_2}) | \Psi^S_{1/2} \Psi^J_{-1/2} \rangle = \\ \label{Z} 
    = \frac{2\sqrt{2}}{\sqrt{3}} \ae |f(0)|^2, \\
\nonumber
    \displaystyle \mathcal{V} = \langle \Psi^S_{1/2} \Psi^J_{-1/2} | U ({\bm r_1} - {\bm r_2}) | \Psi^S_{-1/2} \Psi^J_{1/2} \rangle = \\ \label{V} 
    = 2 \ae |f(0)|^2.
\end{gather}
Here the Coloumb potential is used, which is given by the expression
\begin{equation}
\label{Coloumb_potential}
    \displaystyle U({\bm r_1} - {\bm r_2}) = \frac{e^2}{|{\bm r_1} - {\bm r_2}|}.
\end{equation}
Note that we treat the Coloumb interaction between the localized hole and holes in the $d$-shell (as empty states in the half-filled shell), and hence we have the positive sign in Eq.~(\ref{Coloumb_potential}). The $W$ terms in Eqs.~(\ref{X}--\ref{Y}) could be excluded from the consideration because they result in equal general energy shift of all $4$ states due to the Coloumb interaction. The main result is the connection of $\mathcal{X}$, $\mathcal{Y}$, $\mathcal{Z}$ and $\mathcal{V}$ terms with the exchange integral $\ae$, which reads as
\begin{gather}
\label{J_exchange_integral}
    \displaystyle \ae = \! \sum\limits_{j=1}^{5} 
    \! \iint_\Omega \! \! \frac{{\rm d}{\bm r_1} {\rm d}{\bm r_2}}{15} \varphi^{j*}({\bm r_1}) \varphi^{j}({\bm r_2}) U({\bm r_1} - {\bm r_2}) \chi^*({\bm r_2}) \chi({\bm r_1}), 
\end{gather}
where integrations goes over directly doubled GaAs-crystal elementary cell volume $\Omega$, 
the sum is taken over all five one-electron orbitals of the $3d$-shell of the manganese ion $\varphi^{j}$ (the upper index numerates all possible orbital states $j=1,\dots,5$), and there is an overlapping with a $p$-like Bloch part of the localized hole wave function $\chi$. 

The eigenvalues of matrix (\ref{Coloumb_Hamiltonian_4x4}) give us the following energy differences between eigenstates of this system
\begin{gather}
\nonumber
      E_2 - E_1 = \frac{4}{3}\ae |f(0)|^2, \quad E_3 - E_2 = 2\ae |f(0)|^2, \\ E_4 - E_3 = \frac{8}{3}\ae |f(0)|^2.
\label{Delta_E}
\end{gather}
One can see from (\ref{energy_diff}) that they give the same ratio between energy differences as in the phenomenological approach using Hamiltonian (\ref{H_exchange}). And these results totally coincide if one puts
\begin{equation}
\label{A_expr}
      A = \frac{2}{3}\ae |f(0)|^2.
\end{equation}
The latter expression gives us the tool for microscopic calculations of external forces effects on the exchange constant $A$, which is relevant for a lot of measurements.


\subsection{Exchange constant dependence on deformation.}
From the symmetry point of view possible dependence of exchange constant on deformation
reads as
\begin{equation}
    \label{Hexch_symm}
    \hat{H} = A_0 ( \hat{\bm S} \cdot \hat{\bm J} ) + B_P {\rm Tr} (\hat{\varepsilon}) ( \hat{\bm S} \cdot \hat{\bm J} ) + C_P \sum\limits_{i,j} \hat{S}_i \hat{J}_j \varepsilon_{ij}.
\end{equation}
If one consider hydrostatic deformation, constant $A$ depends only on the trace of deformation
tensor $\varepsilon_{ij} = \delta_{ij} {\rm Tr} (\hat{\varepsilon})/3$ (here $\delta_{ij}$ is the Kronecker delta-symbol)
\begin{equation}
    \label{Hexch_symm_hydrostatic_case}
    \hat{H} = A_0 ( \hat{\bm S} \cdot \hat{\bm J} ) + (B_P + C_P /3) {\rm Tr} (\hat{\varepsilon}) ( \hat{\bm S} \cdot \hat{\bm J} ) .
\end{equation}
Further, we will neglect the dependence of envelope wave functions $f(0)$ on deformation $\varepsilon$,
because their change is too small (it is in the order of $1$~\% of observed values \cite{nestoklon2015fine}).

To understand the microscopic foundations of such Hamiltonian dependence on deformation, we assume that the true wave functions of the $p$-type forming the Bloch eigenstates of the valley band could be admixed by some other atomic states, for example, via the $pd$ hybridization mechanism keeping the total symmetry of the state unchanged. Such hybridization can occur due to different reasons, for example, due to the lack of inversion symmetry in the $T_d$ group or the action of some internal potentials. We suggest here to consider the admixing mechanism stemming from the existence of random electric fields that commonly present near Mn impurity centers in GaAs \cite{averkiev2018mnga}. Such random fields are usually considered as an additional source of fine structure splittings in the Mn acceptor energy spectrum \cite{averkiev2018mnga,Krainov2016}, but they also could affect local wave functions of bounded holes, i.e., the Bloch wave functions due to the $pd$ hybridization. Thus, in Eq.~(\ref{J_exchange_integral}) the $\chi$ functions should be substituted by the hybridized combinations like
\begin{gather}
    \label{pd-hybrid}
    \displaystyle \tilde{\chi} \approx \chi + \sum\limits_d \gamma_d \varphi^d, \quad
    \tilde{\varphi}^d \approx \varphi^d - \gamma_d^* \chi, \quad
    \gamma_d = \frac{\langle \varphi^d | \hat{V} | \chi \rangle }{E_p - E_d} .
\end{gather}
Here $E_p$ and $E_d$ represent the pure atomic energies of pure $p$- and $d$-states without hybridization $E_p - E_d \sim 1$~eV 
(we assume here that for Mn-acceptor in GaAs, the pure d-state is lying not very far from the top of the valence band, and hence $pd$ interaction is the most large one),
and term $\hat{V}= \bm r \cdot \bm F$ stands for the hybridization operator admixing one state to another via electro-dipole induction mechanism, which is due to some local random electric force $\bm F$. The latter could be very sensible to the change of the elementary cell if the deformation of the crystal occurs
\begin{equation}
    \label{Felectricfield}
    F'_i = F_i + \alpha \varepsilon_{ij} F_j ,
\end{equation}
Here we introduce dimensionless parameter $\alpha$ that taking into account deformation dependence of random fields. We assume that the applied stress is a small parameter of the theory, so $\alpha \varepsilon \ll 1$, and further we take into account only linear terms on stress. 

Thus, we can estimate the change of $\ae$ under a pressure or a temperature-affected widening
using the following assumptions about local electric force properties
\begin{gather}
    \nonumber
    \hat{V}^2 \approx r_i r_k (F_i F_k + \alpha \varepsilon_{km} F_i F_m +\alpha \varepsilon_{ij} F_j F_k). \\ \nonumber
    \langle\langle F_i \rangle\rangle = 0, \qquad \langle\langle F_i F_j \rangle\rangle = \zeta \delta_{ij}.
\end{gather}
Here double angle brackets represent averaging by possible realizations of random forces. Of
course, the true averaging should be processed over observable values, although the mean
value of an observable depends on deformation approximately the same way as the observable
calculated with such averaged value of exchange constant.

Finally, one can conclude that after averaging by random forces Eq.~(\ref{J_exchange_integral}) could be represented by the following terms
\begin{equation}
    \label{ae_with_hybrid}
    \ae \approx \ae_{ppdd} + \! \sum_{l,i} \ae_{dddd}^{ll} \frac{\langle \chi | r_i | \varphi^l \rangle \langle \varphi^l | r_i | \chi \rangle}{(E_p - E_d)^2} \zeta \! \left( \! 1 \! + \! \frac{2}{3}\alpha {\rm Tr} (\hat{\varepsilon}) \! \right) ,
\end{equation}
where
\begin{gather}
    \nonumber 
    \ae_{ppdd} = 
    \sum\limits_{j=1}^{5} \int \!\!\! \int_{\Omega} \frac{{\rm d} r_1 {\rm d} r_2}{15} \varphi^{j*} (\bm r_1) \varphi^j (\bm r_2) U({\bm r_1 - \bm r_2 }) \times \\ \nonumber
    \times \chi^* (\bm r_2) \chi (\bm r_1), \\
    \nonumber
    \ae_{dddd}^{ln} = 
    \sum\limits_{j=1}^{5} \int \!\!\! \int_{\Omega} \frac{{\rm d} r_1 {\rm d} r_2}{15} \varphi^{j*} (\bm r_1) \varphi^j (\bm r_2) U({\bm r_1 - \bm r_2 }) \times \\ \nonumber
    \times \varphi^{l*} (\bm r_2) \varphi^{n} (\bm r_1),
\end{gather}
which are exchange integrals with different integrand functions. We should note that the values of these terms depend on the functions overlap, and hence the more $d$-functions of Mn ion are involved, the larger the value of the Coulomb term is
$\ae_{ppdd} \ll \ae_{dddd}$.

To estimate the magnitude of the effect, we first take into account that all exchange integrals between $d$-functions have the same value in sum in eq.~\ref{ae_with_hybrid}. Then using the hydrogen atom functions $\chi$ corresponding to 4$p$ orbitals and $\varphi^d$ corresponding to 3$d$ orbitals one can obtain an estimate $\ae_{dddd}/\ae_{ppdd} \approx 10^4$. Also, we can take matrix elements of coordinates approximately equal to the Bohr radius
of the atom $\langle \chi |r_i | \varphi^l \rangle \approx \langle \varphi^l | r_i | \chi \rangle \approx a_B \approx 10^{-8}$~cm, and the value of the random forces dispersion could be estimated as having the order of a typical interatomic interaction term 
$\sqrt{\zeta} = F_* \approx 10^6$~eV/cm (which is comparable with typical values of the mean force affecting the nuclear complex of the lattice cell in GaAs in the case of the Cu ion, for which $F \approx 5 \cdot 10^6$~eV/cm \cite{baryshnikov2012resonant,averkiev2014ultrasonic}). Then we can write an estimate for exchange constant $A$ change with deformation ($A_P \equiv  B_P + C_P/3$)
\begin{equation}
    \label{APexpr}
    A = A_0 + A_P {\rm Tr} (\hat{\varepsilon}),
\end{equation}
where
\begin{equation}
    \label{APestimate}
    \frac{A_P}{A_0} \approx \frac{2}{3} \alpha \frac{(15(a_BF_*)^2/(E_p - E_d)^2) \ae_{dddd}/\ae_{ppdd}}{1 + (15(a_BF_*)^2/(E_p - E_d)^2) \ae_{dddd}/\ae_{ppdd}}. 
\end{equation}
From data analysis in \cite{Krainov2016}, we can estimate alpha as $A_P / A_0 = (900~{\rm meV}/2.5~{\rm meV}) \approx 360$, which is equivalent to the relative change of $A$ nearly by $ \delta A / A_0 \sim \alpha {\rm Tr} (\varepsilon) \sim 0.2$ at a half of critical strain of GaAs crystal corresponding to hardness limit at helium temperatures.



\section{Calculations and Discussion}
We have discussed above the parametric dependence of exchange constant value on crystal deformation and its microscopic reasons.
This fact had already played its role in the explanation of Raman scattering experiment results \cite{Krainov2016}, and now we are going to demonstrate clearly that the same fact is responsible for 
high temperature magnetic susceptibility reduction 
measured independently in a completely different experimental setting \cite{averkiev2018mnga}. 

\begin{figure}[h!]
\includegraphics[scale=0.9]{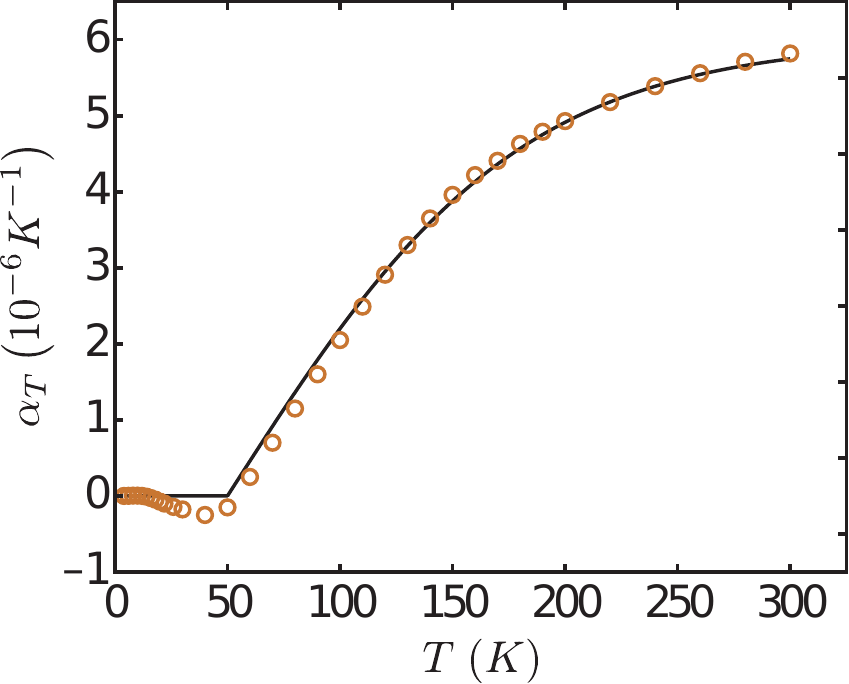}
\caption{\label{alphaT} Temperature dependence of linear expansion coefficient $\alpha_T$. Dark orange circles are experimental results from \cite{Novikova74} (see Table 80 on page 233), black line is our interpolation for this dependence up to $300$~K.}
\end{figure}

As GaAs crystal undergoes thermal expansion, we are going to test our hypothesis of this expansion being responsible for anomalous reduction of magnetic susceptibility at relatively high temperatures. The temperature dependence of linear expansion coefficient $\alpha_T$ could be found in literature (see, for example, \cite{IoffeDataBase} or \cite{Novikova74}). We show this dependence in Fig.~\ref{alphaT}.

We will use a simple function to interpolate the $\alpha_T$ dependence on temperature, which makes the interpolation work up to $300$~K quite well (see Fig.~\ref{alphaT})
\begin{equation}
    \label{alphaT_interpolation}
    \displaystyle \tilde{\alpha}_T = 
    \left\{
        \begin{array}{ll}
            0, & T<50\!~K,  \\
            \displaystyle C \tanh{\left(\frac{T - 50}{180 - 50}\right)}, & 50\!~K \leq T \leq 300\!~K.
        \end{array}
    \right.
\end{equation}
It is implied that $T$ is measured in kelvins. The coefficient $C=6\cdot 10^{-6}~K^{-1}$. Note that there is a slight increase in the $\alpha_T$ coefficient above $300$~K (at $800$~K it reaches $7.4\cdot 10^{-6}$~K$^{-1}$, see the full table of its values in \cite{Novikova74}), and hence the approximation in Eq.~(\ref{alphaT_interpolation}) does not work if $T>300$~K. But for our purposes it is enough to consider the region of $T<300$~K, in which the interpolation in Eq.~(\ref{alphaT_interpolation}) describes experimental data quite well. Note that a very small decrease in $\alpha_T$ values between $25$~K and $50$~K does not affect the observables in any reasonable manner, thus, we neglect it. 

\begin{figure}[h!]
\includegraphics[scale=0.65]{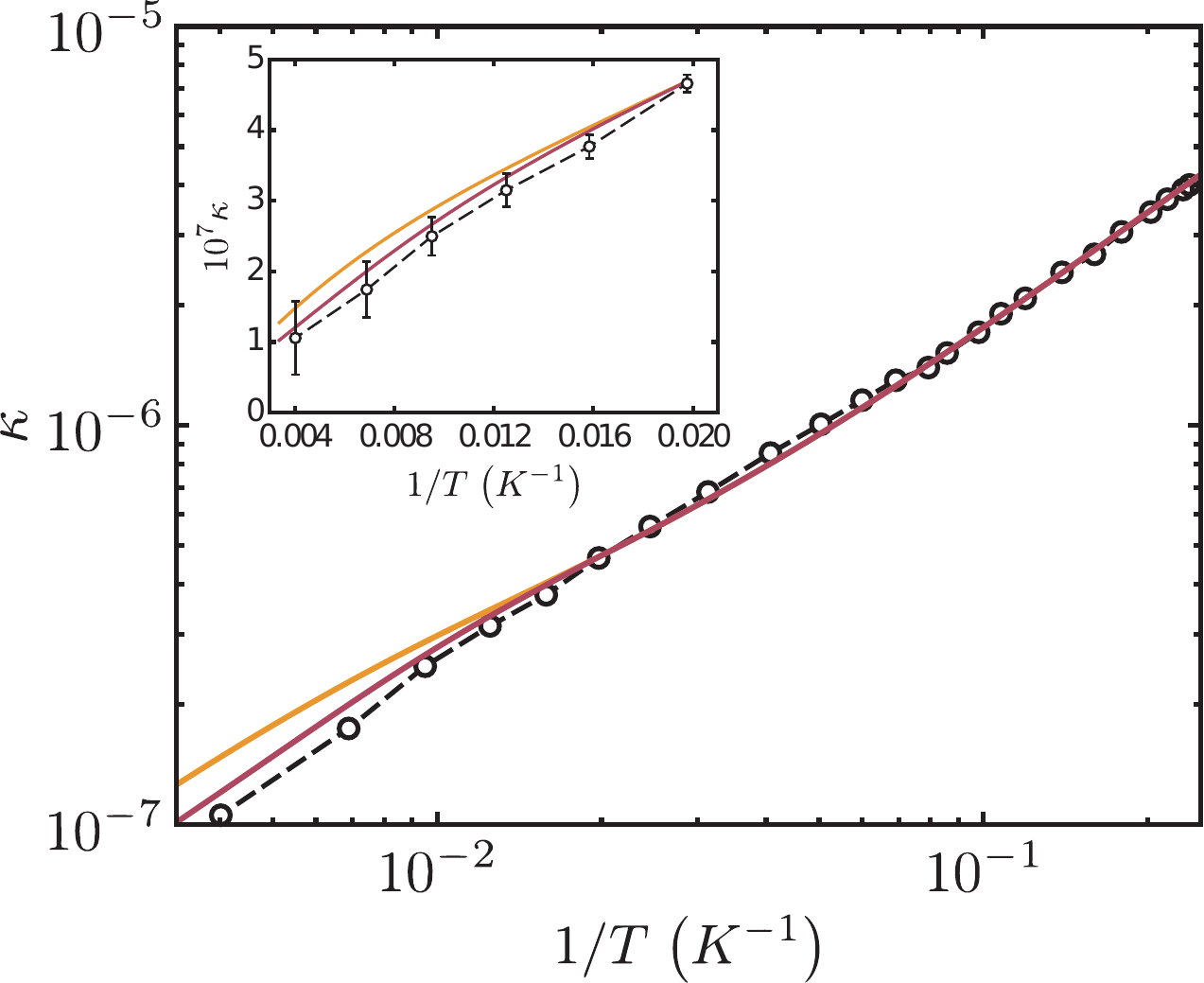}
\caption{\label{magsus} Temperature dependence of static magnetic susceptibility $\kappa$ of manganese ions in GaAs crystal (concentration of manganese ions $5.3\cdot 10^{18}$~cm$^{-3}$). Both axes have logarithmic scale. Black circles connected by dash lines represent experimental data from \cite{averkiev2018mnga,andrianov1983characteristics}. Light orange solid line is a result of calculation based on the ordinary theory from \cite{averkiev2018mnga}, which implies that $A(T) = A_0 = 2.6$~meV. Violet solid line represents our calculation result, where the expression for magnetic susceptibility taken from \cite{averkiev2018mnga} is modified by taking into account the exchange constant variation with temperature $A(T)$ due to thermal expansion effect.}
\end{figure}

We are interested in temperature range $T = 0 \div 300$~K. So we can write the dependence of exchange value $A$ on $T$ taking into account Eq.~(\ref{APexpr})
\begin{equation}
    \label{A_dep_on_T}
    A(T) = A_0 + A_P \cdot 3\tilde{\alpha}_T \cdot T,
\end{equation}
where $A_0 = 2.6$~meV \cite{Krainov2016}. We have multiplied $\tilde{\alpha}_T$ by a factor of $3$ to get the bulk thermal expansion coefficient from the linear one, because ${\rm Tr} (\varepsilon) = \varepsilon_{xx}(T) + \varepsilon_{yy}(T) + \varepsilon_{zz}(T) = 3 \varepsilon_{xx}(T)$. Here we use the same value of $A_P = 900$~meV as in \cite{Krainov2016}. One can see the calculations results in Fig.~\ref{magsus}. Note that in \cite{averkiev2018mnga} the electron-hole basis is used, hence one should change the sign of the exchange constant into opposite one compared with the our result to obtain the same order of energy levels for manganese acceptor. Thus, substituting Eq.~\ref{A_dep_on_T} into the formulas in \cite{averkiev2018mnga}, we need to multiply $A(T)$ by $(-1)$.

As can be seen from Fig.~\ref{magsus} the relative mismatch between theory and experimental results at $T>100$~K reduces approximately from $50$\% to $20$\%, if we use Eq.~\ref{A_dep_on_T}.
This reduction of the systematic mismatch leads to a better agreement between theoretical results and the experimental data in the high-temperature region, which have the allowable magnitude of the experimental error (see discussion in \cite{averkiev2018mnga}, experimental data have been first obtained in \cite{andrianov1983characteristics}, and the same mismatch has also been independently mentioned in \cite{frey1988paramagnetism}). 
Also we point out that the sign of changes of  exchange interaction constant, which we use to fit magnetic susceptibility data, is the same as used in Raman experiments \cite{Krainov2016}. 
Note that other possible factors, such as crystal field effect or reduction of magnetic susceptibility caused by the dynamical Jahn-Teller effect observed by us in Supplementary materials, give no pronounce effects on magnetic susceptibility. Moreover, their effects diminish at high temperatures, and they also ruin the well-established theory predictions at low temperatures below $50$~K.

Thus, the effect of exchange constant parametric dependence on lattice deformation is the only effect that provides reasonable explanation of both high and low temperature behaviour characteristics of the manganese acceptor center in GaAs. Note also that $A(T)$ at $T=300$~K is nearly three times larger than $A_0$, and it could reach even higher values at higher temperatures according to \cite{Novikova74} and Eq.~\ref{A_dep_on_T}. 
Note that at such big changes in $A$ the nonlinear terms on lattice deformation should be also taken into account in the $pd$ hybridization mechanism of exchange constant renormalization via random fields as soon as the parameter $\alpha {\rm Tr} (\varepsilon)$ reaches and exceeds the limit of $1$. But we show in Fig.~\ref{magsus} that even linear terms give the right trend in temperature dependence of magnetic susceptibility.


\section{Conclusion}
Exchange constant value between $d$-electrons of manganese ion impurity in GaAs crystal and the hole, localized from the valence band on the impurity ion, is microscopically derived. The effect of crystal lattice period change on the value of the exchange coupling constant occurring via the hybridization of exchanging orbitals is shown and estimated. We also discuss the effect of the thermal expansion causing the change in magnetic susceptibility. 
We show that accounting for this effect leads to a better agreement between theoretical results and magnetic susceptibility data measured at high temperatures. The considered thermal widening mechanism does not influence the low-temperature magnetic susceptibility behaviour. This result is also in agreement with another experiment of Raman scattering on Mn acceptors in GaAs with applied external strain. 
We believe that the approach to the analytical calculation of the exchange constant could be generalized to the case of 
other magnetic centers in semiconductor structures and semi-magnetic compounds.


\begin{acknowledgments}
This work has been supported by the Russian Science Foundation (analytical theory -- Project  18-72-10111).
K.~A.~B. thanks the Theoretical Physics and Mathematics Advancement Foundation "BASIS". We also thank M.~O. Nestoklon, S.~A. Tarasenko, and M.~M. Glazov for fruitful discussions.
We dedicate this article to the memory of our colleague and co-author V.~F. Sapega (Ioffe Institute) who passed away in 2022.
\end{acknowledgments}


\appendix

\section{Calculation of exchange integrals.}
\label{App1}
The localized-on-the-ion hole has the Bloch part of wave function, which describes both spin and orbital degrees of freedom in $\Gamma_8$-state
\begin{eqnarray}
\label{Bloch_G8}
      \Psi^J_{3/2} = - \alpha \frac{X+iY}{\sqrt{2}} f(0); \\
      \Psi^J_{1/2} = \left( \sqrt{\frac{2}{3}} \alpha Z - \beta \frac{X+iY}{\sqrt{6}} \right) f(0); \\
      \Psi^J_{-1/2} = \left( \sqrt{\frac{2}{3}} \beta Z + \alpha \frac{X-iY}{\sqrt{6}} \right) f(0); \\
      \Psi^J_{-3/2} = \beta \frac{X-iY}{\sqrt{2}} f(0).
\end{eqnarray}
Here $\alpha$ and $\beta$ means spin-up and spin-down states of the hole captured and localized from the valley band of GaAs crystal, respectively. Space orbitals $X$, $Y$ and $Z$ correspond to a $p$-like orbitals, which form the valley band of the crystal, and hence one can prove that they are quite similar from the cubic symmetry point of view. So we will use more compact notations as $\chi^{+} = -(X+iY)/\sqrt{2}$ and $\chi^{-}=(X-iY)/\sqrt{2}$. 

The half-filled $3d$-shell of the Mn ion is described by a five-hole wave function with the totally symmetrical spin part. We assume that Hund's rule is the most powerful here, and all spin-spin interaction in the shell has already led to the appearance of co-directed spins of all five $d$-holes resulting in the total spin $S=5/2$, and hence it has an antisymmetric orbital part
\begin{gather}
\label{d_shell}
      \Psi^S_{S_z} = \Phi^d_{1,2,3,4,5} | S, S_z \rangle; \\
      \Phi^d_{1,2,3,4,5} = \frac{1}{\sqrt{5!}}  
      \begin{vmatrix}
          \varphi^1_1 & \varphi^2_1 & \varphi^3_1 & \varphi^4_1 & \varphi^5_1 \\
          \varphi^1_2 & \varphi^2_2 & \varphi^3_2 & \varphi^4_2 & \varphi^5_2 \\
          \varphi^1_3 & \varphi^2_3 & \varphi^3_3 & \varphi^4_3 & \varphi^5_3 \\
          \varphi^1_4 & \varphi^2_4 & \varphi^3_4 & \varphi^4_4 & \varphi^5_4 \\
          \varphi^1_5 & \varphi^2_5 & \varphi^3_5 & \varphi^4_5 & \varphi^5_5 \\
      \end{vmatrix} ; \label{orbital_functions} \\
      | S, 5/2 \rangle = \Theta^{5/2}_{1,2,3,4,5} = \alpha_1 \alpha_2 \alpha_3 \alpha_4 \alpha_5 ; \label{S5_2} \\
      | S, 3/2 \rangle = \Theta^{3/2}_{1,2,3,4,5} = \nonumber \\
      = \frac{1}{\sqrt{5}} \left(\alpha_1 \alpha_2 \alpha_3 \alpha_4 \beta_5 + \alpha_1 \alpha_2 \alpha_3 \beta_4 \alpha_5 + \dots \right); \label{S3_2}\\
      | S, 1/2 \rangle = \Theta^{1/2}_{1,2,3,4,5} = \frac{1}{\sqrt{10}} \left(\alpha_1 \alpha_2 \alpha_3 \beta_4 \beta_5 + \right. \nonumber \\
      \left. + \alpha_1 \alpha_2 \beta_3 \alpha_4 \beta_5 + \dots + \alpha_1 \alpha_2 \beta_3 \beta_4 \alpha_5 + \dots \right). \label{S1_2}
\end{gather}
The lower indices of the $d$-holes orbital coordinates ${\bm r}_1,{\bm r}_2,{\bm r}_3,{\bm r}_4,{\bm r}_5$ are the lower indices $k=1,2,3,4,5$ of the functions in Eq.~(\ref{orbital_functions}).
The upper indices of $\varphi^j_k$ functions list five $d$-shell different orbitals $j=1,2,3,4,5$.
According to Hund’s rule we take all the five possible $d$-orbitals for the ground state of the ion, because the states with identical orbital functions (and hence with opposite directions of spins) correspond to the excited states of the Mn ion having the excitation energy of electronvolts, and they are out of consideration. 
Spin coordinates of different $d$-holes are also indicated by the corresponding indices.
The dots in the brackets of (\ref{S3_2}) and (\ref{S1_2}) mean that all possible permutations of four $\alpha$ and one $\beta$ for (\ref{S3_2}) and three $\alpha$ and two $\beta$ for (\ref{S1_2}) over $d$-hole indices are taken into account. Wave functions corresponding to the negative projections of the total spin on the $z$ axis $S_z = -1/2,-3/2,-5/2$ are the same if one changes all $\alpha$ to $\beta$ and vice versa. The normalization constants for those wave functions are equal to one over square root of the number of permutations of $\alpha$ and $\beta$ positions in each case, i.e., $1/\sqrt{C_5^2} = 1/\sqrt{10}$, $1/\sqrt{C_5^1} = 1/\sqrt{5}$, $1/\sqrt{C_5^0} = 1$, respectively. 

Thus, we have five $d$-holes in the $3d$-shell, where the strongest Coloumb interaction has already led to realization of Hund's rule, and there is the sixth localized-on-the-ion hole in $\Gamma_8$ state, which interacts with all those five $d$-holes. Let us introduce the potential energy operator of remaining weaker Coloumb interactions between the particles
\begin{equation}
\label{Coloumb_sum}
      \displaystyle \hat{U} = \sum\limits_{i=1}^{5} U({\bm r_i} - {\bm r_6}) .
\end{equation}
Let us calculate the first diagonal element of such Coloumb operator in the basis of zero total-momentum projection functions. Using the notation introduced above, we can write an antisymmetrized form of the wave function
\begin{gather}
\label{PsiS3/2_PsiJ-3/2}
    \Psi^S_{3/2} \Psi^J_{-3/2} = \frac{f(0)}{\sqrt{6}} \left\{ \Phi^d_{1,2,3,4,5} \Theta^{3/2}_{1,2,3,4,5} ~\! \chi^{-}_6 \beta_6 - \right. \nonumber \\ 
    \left. - \Phi^d_{1,2,3,4,6} \Theta^{3/2}_{1,2,3,4,6} ~\! \chi^{-}_5 \beta_5 - \right. \nonumber \\ 
    \left. - \Phi^d_{1,2,3,6,5} \Theta^{3/2}_{1,2,3,6,5} ~\! \chi^{-}_4 \beta_4 - \dots \right\}. 
\end{gather}
This many-particle wave function is formed by the multiplication of wave functions of the localized hole and of five $d$-holes with the fixed order of their coordinates ($i=1,2,3,4,5$), followed by subtraction of all possible multiples with consequently interchanged coordinates of the localized hole ($i=6$) and the $d$-shell holes. One can prove that this procedure gives us the antisymmetric total wave function of the system in accordance with the properties of determinant columns interchange. 

Here we illustrate this result with the example of a three-electron system. If one has an antisymmetric combination of two electron wave functions $\phi_{1,2} = \varphi_1 \psi_2 - \varphi_2 \psi_1$ with the fixed order of arguments, then one can show that the procedure gives us the fully antisymmetric wave function when adding the third electron
\begin{gather}
\label{as_procedure}
    \displaystyle \phi_{1,2} \chi_3 - \phi_{1,3} \chi_2 - \phi_{3,2} \chi_1 = \nonumber \\
    = (\varphi_1 \psi_2 - \varphi_2 \psi_1) \chi_3 - (\varphi_1 \psi_3 - \varphi_3 \psi_1) \chi_2 - \nonumber \\
    - (\varphi_3 \psi_2 - \varphi_2 \psi_3) \chi_1 = \nonumber \\
    = 
    \begin{vmatrix}
          \varphi_1 & \varphi_2 & \varphi_3 \\
          \psi_1 & \psi_2 & \psi_3 \\
          \chi_1 & \chi_2 & \chi_3 \\
      \end{vmatrix} .
\end{gather}

Then
\begin{gather}
\label{X_calc}
    \displaystyle \mathcal{X} = \langle \Psi^S_{3/2} \Psi^J_{-3/2} | U ({\bm r_1} - {\bm r_2}) | \Psi^S_{3/2} \Psi^J_{-3/2} \rangle = \nonumber \\
    = \frac{|f(0)|^2}{6} \int {\rm d}{\bm r_1} \dots {\rm d}{\bm r_6} \times \nonumber \\
    \times \left( \Phi^{d*}_{1,2,3,4,5}  ~\! \Theta^{3/2 \dagger}_{1,2,3,4,5}  ~\! \chi^*_6  ~\! \beta^{\dagger}_6 - (5)^{\dagger} - (4)^{\dagger} - \dots \right) \times \nonumber \\
    \times \hat{U} \times \left( \Phi^{d}_{1,2,3,4,5}  ~\! \Theta^{3/2}_{1,2,3,4,5}  ~\! \chi_6  ~\! \beta_6 - (5) - (4) - \dots \right) = \nonumber \\
    = W + \ae |f(0)|^2,
\end{gather}
where the notation $(5)$, $(4)$ and etc. are introduced for the terms, in which the localized hole index $6$ (and hence its coordinate ${\bm r}_6$), is swapped with the corresponding intershell $d$-hole index $5$, $4$ and etc.

The Coloumb term $W$ is determined by the direct product of the multiples of the same type as it is shown on the scheme in Fig~\ref{direct} below, and it reads as
\begin{equation}
\label{W}
    \displaystyle W = |f(0)|^2 \iint {\rm d}{\bm r_1}{\rm d}{\bm r_2} 
    | \chi^{-} ({\bm r_1}) |^2 \sum\limits_{j=1}^5 | \varphi^j ({\bm r_2}) |^2 U({\bm r_1} - {\bm r_2}).
\end{equation}

\begin{figure}[h!]
\includegraphics[scale=0.2]{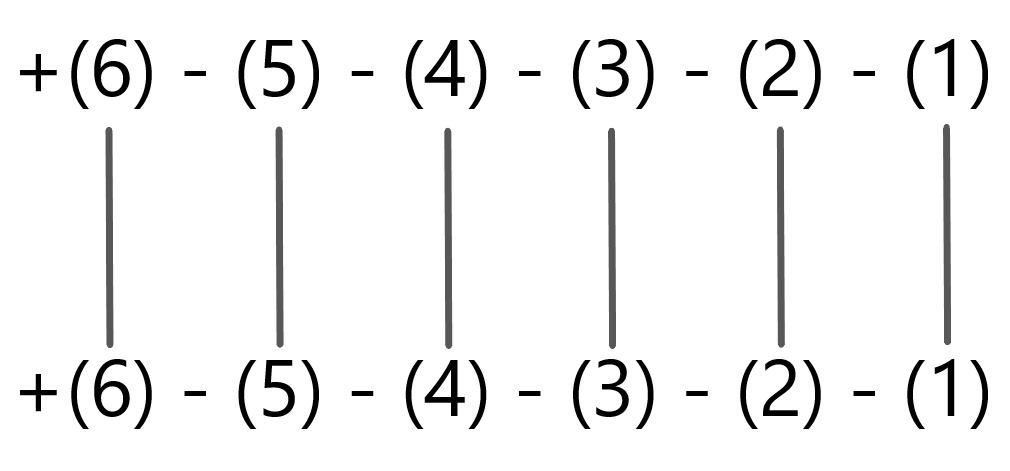}
\caption{\label{direct} The scheme of bra and ket direct multiples in Eq.~\ref{X_calc}. The total number of direct multiples is equal to $6$.}
\end{figure}

\begin{figure}[h!]
{\includegraphics[scale=0.2]{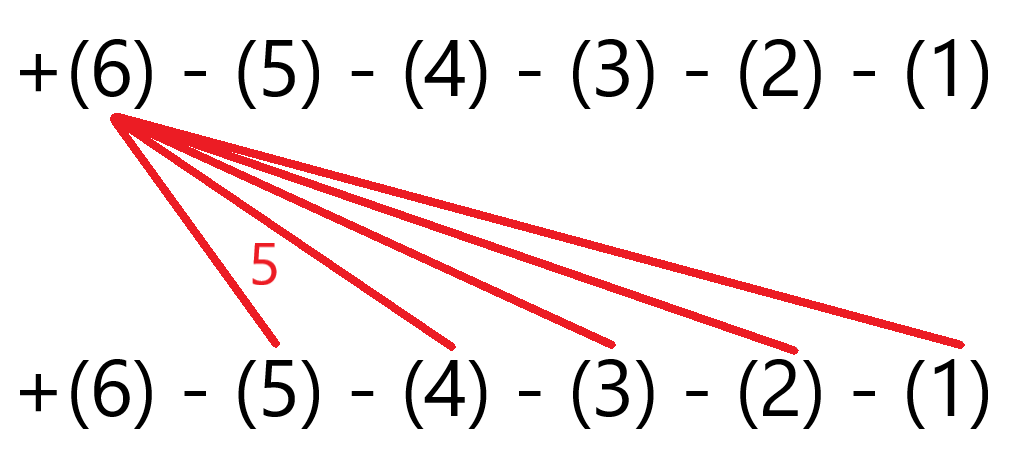} (a)}
\quad
{\includegraphics[scale=0.2]{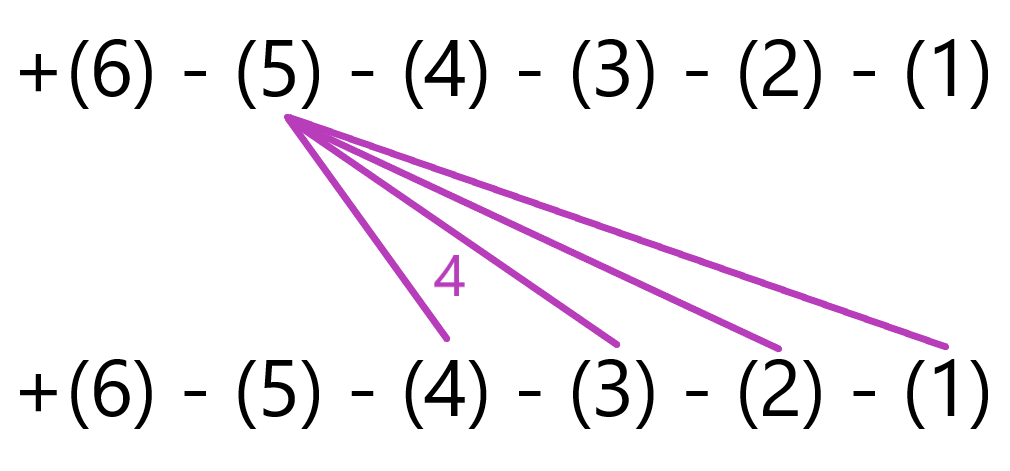} (b)}
\quad
{\includegraphics[scale=0.2]{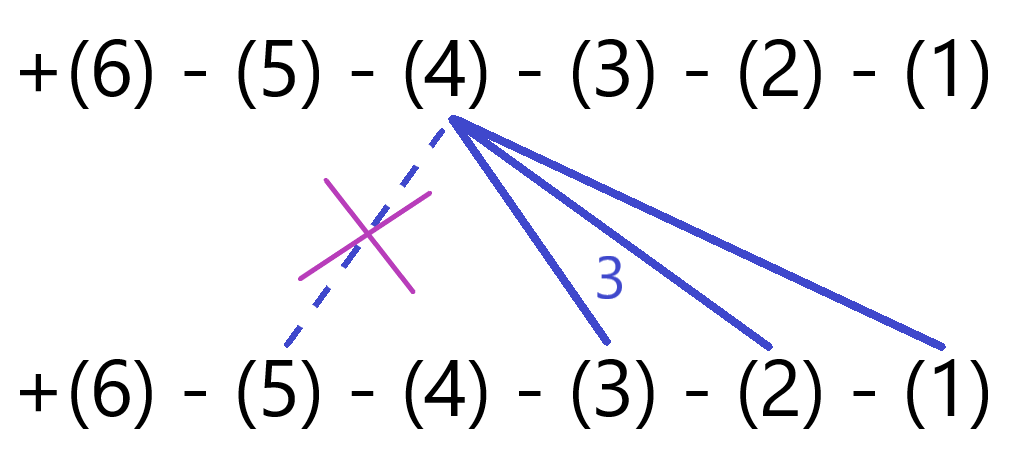} (c)}
\quad
{\includegraphics[scale=0.2]{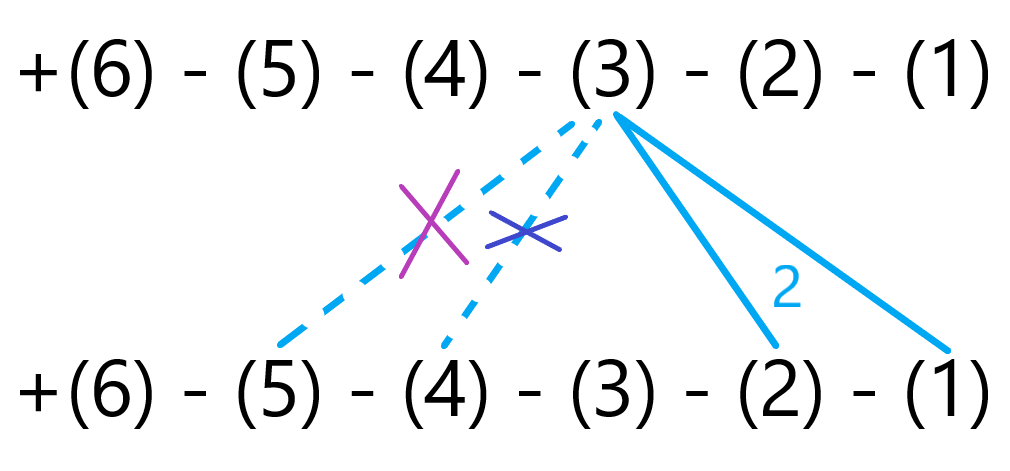} (d)}
\quad
{\includegraphics[scale=0.2]{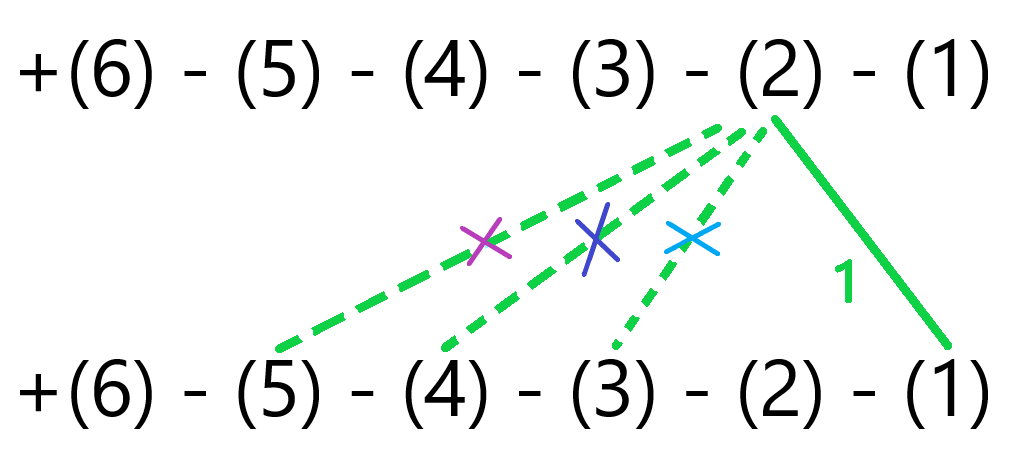} (e)}
\caption{\label{multiples} The scheme of bra and ket cross-multiples in Eq.~(\ref{X_calc}). (a): There are $5 \cdot 2 = 10$ cross-multiples with the minus sign if one consider five cross-multiples in the inset and their mirror twins emerging as if the virtual reflection in the horizontal plane takes place. (b), (c), (d) and (e): The number of each multiple of the plus sign should be multiplied by $2$ due to the same reason as discussed for inset (a). The total number of multiples equals to $(4+3+2+1)\cdot 2 = 20$.}
\end{figure}

And the exchange term is given by Eq.~(\ref{J_exchange_integral}), where $\chi$ stands for $\chi^{-}$ and the numerical prefactor stems from the normalization of wave functions and convolution of spin wave functions with all possible cross-multiples with permutable indices, which are shown in Figs~\ref{multiples}(a) -- \ref{multiples}(e). Note that all multiples give the same contribution but with different signs. Thus, we carry out the calculations for the case of multiplication of $(6)$ and $(5)$ terms shown in Fig.~\ref{multiples}(a) and take proper account of the summation of all terms with positive and negative signs
\begin{equation}
\label{first_term_in_5}
    \displaystyle \Theta^{3/2 \dagger}_{1,2,3,4,5} ~\! \beta^{\dagger}_6  ~\! \Theta^{3/2}_{1,2,3,4,6}  ~\! \beta_5  \left( -5 \cdot 2 + (4+3+2+1) \cdot 2 \right) = 2.
\end{equation}


Note also that the remaining orbital part of the exchange integral (after the summation by the spin indices) has the following form
\begin{gather}
\label{ae2}
    \displaystyle \ae =
    \frac{2}{6} \int {\rm d}{\bm r_1} \dots {\rm d}{\bm r_6} ~\! \chi^*_6  ~\! \chi_5  ~\! \Phi^{d*}_{1,2,3,4,5}  ~\! \Phi^{d}_{1,2,3,4,6} ~\! \times \nonumber \\
    \times \left( U(|{\bm r_5} - {\bm r_6}|) + \dots \right) = \nonumber \\
     =
    \frac{1}{3} \cdot \frac{4!}{5!} \iint {\rm d}{\bm r_5} {\rm d}{\bm r_6} ~\! \chi^*_6  ~\! \chi_5  \sum\limits_{j=1}^{5} \varphi^{j*}_{5}  \varphi^{j}_{6} ~\! U(|{\bm r_5} - {\bm r_6}|).
\end{gather}
The integration with all terms denoted by dots in the first part of Eq.~(\ref{ae2}) gives us zero due to the orthogonality of all orbital wave functions. The summation over $d$-orbital indices in the last part of the equation is carried out considering only one index $j=1,\dots 5$ for both one-particle functions $\varphi^{j*}_{5}$ and $\varphi^{j}_{6}$. The latter could be easily checked by treating the multiples in explicit forms written one under another
\begin{gather}
\label{explicit_multiples1}
    \displaystyle \chi^*_6  ~\! \Phi^{d*}_{1,2,3,4,5} = \frac{\chi^*_6}{\sqrt{5!}} \left(  \varphi^1_1 \varphi^2_2 \varphi^3_3 \varphi^4_4 \varphi^5_5 - \varphi^1_1 \varphi^2_2 \varphi^3_3 \varphi^4_5 \varphi^5_4 + \dots \right)^* \! , \\
\label{explicit_multiples2}
    \chi_5  ~\! \Phi^{d}_{1,2,3,4,6} = \frac{\chi_5}{\sqrt{5!}} \left(  \varphi^1_1 \varphi^2_2 \varphi^3_3 \varphi^4_4 \varphi^5_6 - \varphi^1_1 \varphi^2_2 \varphi^3_3 \varphi^4_6 \varphi^5_4 + \dots \right).
\end{gather}
Here the first term is determined by the fixed sequence of the coordinate indices, and all others are determined by the coordinate indices swapping accompanied by a change of the sign.
One can see that the non-zero multiples are only those which are the products of two terms written strictly under each other in Eqs.~(\ref{explicit_multiples1}) and (\ref{explicit_multiples2}).
All other multiples gives us zero due to the mutual orthogonality of all functions. 
Thus, all terms are summed up only with positive signs.
The number of such summands with fixed position of $5$-th and $6$-th particles equals to the number of permutation of other four electrons over remaining four orbitals, and hence it is equal to $4!=24$.

Let us explain now, in brief, the calculation details for the $\mathcal{Y}$,  $\mathcal{Z}$ and $\mathcal{V}$ terms. The $\mathcal{Y}$ term also involves bra and ket functions of the same type
\begin{gather}
    \Psi^S_{1/2} \Psi^J_{-1/2} = \nonumber \\
    = \frac{f(0)}{\sqrt{6}} \left\{ \Phi^d_{1,2,3,4,5} \Theta^{1/2}_{1,2,3,4,5} ~\! \left( \sqrt{\frac{2}{3}} Z_6 \beta_6 + \sqrt{\frac{1}{3}} \chi^{-}_6 \alpha_6 \right) - \right. \nonumber \\ 
    \displaystyle \left. - (5) - (4) - \dots \right\}.
    \label{PsiS1/2_PsiJ-1/2}
\end{gather}
Here the same notation ($(5)$, $(4)$, etc.) is introduced as for the Eq.~(\ref{X_calc}). 
One can see that the same scheme which we use when calculating the matrix elements of direct Coulomb interaction terms gives us, as denoted in Fig.~\ref{direct}, the same value $W$ (due to the symmetry properties of $\chi^{-}$ and $Z$ functions of the localized hole), and the latter could be excluded from consideration.
The exchange terms calculation requires consideration of two possible results of wave-functions spin parts convolution. 
The first is
\begin{gather}
\label{spin_part_for_Y1}
    \displaystyle \Theta^{1/2 \dagger}_{1,2,3,4,5} ~\! \beta^{\dagger}_6  ~\! \Theta^{1/2}_{1,2,3,4,6}  ~\! \beta_5 = \nonumber \\
    \displaystyle = \frac{1}{10} \left( \alpha_1^{\dagger} \alpha_2^{\dagger} \alpha_3^{\dagger} \beta_4^{\dagger} \beta_5^{\dagger} + \alpha_1^{\dagger} \alpha_2^{\dagger} \beta_3^{\dagger} \alpha_4^{\dagger} \beta_5^{\dagger} + \dots \right) \beta_6^{\dagger} \times \nonumber \\
    \times
    \left( \alpha_1 \alpha_2 \alpha_3 \beta_4 \beta_6 + \alpha_1 \alpha_2 \beta_3 \alpha_4 \beta_6 + \dots \right) \beta_5 = \nonumber \\
    \displaystyle = \frac{1}{10} \left( \alpha_1^{\dagger} \alpha_2^{\dagger} \alpha_3^{\dagger} \beta_4^{\dagger} + \alpha_1^{\dagger} \alpha_2^{\dagger} \beta_3^{\dagger} \alpha_4^{\dagger} + \dots \right)  \times \nonumber \\
    \times
    \left( \alpha_1 \alpha_2 \alpha_3 \beta_4 + \alpha_1 \alpha_2 \beta_3 \alpha_4 + \dots \right) = \frac{4}{10}. 
\end{gather}
and the second is
\begin{gather}
\label{spin_part_for_Y2}
    \displaystyle \Theta^{1/2 \dagger}_{1,2,3,4,5} ~\! \alpha^{\dagger}_6  ~\! \Theta^{1/2}_{1,2,3,4,6}  ~\! \alpha_5 = \nonumber \\
    \displaystyle = \frac{1}{10} \left( \alpha_1^{\dagger} \alpha_2^{\dagger} \alpha_3^{\dagger} \beta_4^{\dagger} \beta_5^{\dagger} + \alpha_1^{\dagger} \alpha_2^{\dagger} \beta_3^{\dagger} \alpha_4^{\dagger} \beta_5^{\dagger} + \dots \right) \alpha_6^{\dagger} \times \nonumber \\
    \times 
    \left( \alpha_1 \alpha_2 \alpha_3 \beta_4 \beta_6 + \alpha_1 \alpha_2 \beta_3 \alpha_4 \beta_6 + \dots \right) \alpha_5 = \nonumber \\
    \displaystyle = \frac{1}{10} \left( \alpha_1^{\dagger} \alpha_2^{\dagger} \beta_3^{\dagger} \beta_4^{\dagger} + \alpha_1^{\dagger} \beta_2^{\dagger} \alpha_3^{\dagger} \beta_4^{\dagger} + \dots \right) \times \nonumber \\
    \times 
    \left( \alpha_1 \alpha_2 \beta_3 \beta_4 + \alpha_1 \beta_2 \alpha_3 \beta_4 + \dots \right) = \frac{6}{10}. 
\end{gather}
Then, after the summation over the spin indices and taking into account possible cross-multiples, as in Figs.~\ref{multiples}(a) -- \ref{multiples}(e), we obtain an additional multiplier $( -5 \cdot 2 + (4+3+2+1) \cdot 2) =10$ as in Eq.~(\ref{first_term_in_5}), and then we get the exchange part of $\mathcal{Y}$ equal to
\begin{gather}
\label{ae_part_Y}
    \displaystyle |f(0)|^2 \frac{10}{6} \int {\rm d}{\bm r_1} \dots {\rm d}{\bm r_6} ~\! \left( \frac{2}{3} \frac{4}{10} Z^*_6 ~\! Z_5 + \frac{1}{3} \frac{6}{10} \chi^*_6  ~\! \chi_5 \right) \times \nonumber \\
    \times 
    \Phi^{d*}_{1,2,3,4,5}  ~\! \Phi^{d}_{1,2,3,4,6} ~\! \left( U(|{\bm r_5} - {\bm r_6}|) + \dots \right) = \nonumber \\
     =
    |f(0)|^2 \frac{7}{9} \cdot \frac{4!}{5!} \iint {\rm d}{\bm r_5} {\rm d}{\bm r_6} ~\! \chi^*_6  ~\! \chi_5 \times \nonumber \\
    \times
    \sum\limits_{j=1}^{5} \varphi^{j*}_{5}  \varphi^{j}_{6} ~\! U(|{\bm r_5} - {\bm r_6}|) = \frac{7}{3} \ae |f(0)|^2 . 
\end{gather}
Here we used the symmetry equivalence of $Z$ and $\chi$ functions when calculating such type of integrals.

When calculating $\mathcal{Z}$, the off-diagonal matrix element between quantum states from Eq.~(\ref{PsiS3/2_PsiJ-3/2}) and Eq.~(\ref{PsiS1/2_PsiJ-1/2}) is taken. Thus, there is no Coloumb term, and the exchange integral in this case reads as 
\begin{gather}
\label{ae_part_Z}
    \displaystyle \mathcal{Z} = |f(0)|^2 \frac{10}{6} \int {\rm d}{\bm r_1} \dots {\rm d}{\bm r_6} ~\! \Phi^{d*}_{1,2,3,4,5} ~\! \Theta_{1,2,3,4,5}^{3/2 \dagger}  ~\! \chi^{-*}_6 ~\! \beta^{\dagger}_6 \times \nonumber \\
    \times
    \hat{U} ~\! \Phi^{d}_{1,2,3,4,6} ~\! \Theta_{1,2,3,4,6}^{1/2} \left( \sqrt{\frac{2}{3}} Z_5 \beta_5 + \sqrt{\frac{1}{3}} \chi^{-}_5 \alpha_5 \right) = \nonumber \\
     =
    |f(0)|^2 \frac{10}{6}  \int {\rm d}{\bm r_1} \dots {\rm d}{\bm r_6} ~\! \Phi^{d*}_{1,2,3,4,5} ~\! \Theta_{1,2,3,4,5}^{3/2 \dagger}  ~\! \chi^{-*}_6 ~\! \beta^{\dagger}_6 \times \nonumber \\
    \times 
    \hat{U} ~\! \Phi^{d}_{1,2,3,4,6} ~\! \Theta_{1,2,3,4,6}^{1/2} \sqrt{\frac{1}{3}} \chi^{-}_5 \alpha_5 . 
\end{gather}
The multiplier $10$ arises as a result of the usage of the introduced scheme, which implies taking into account all exchange integrals (see Figs.~\ref{multiples}(a) -- \ref{multiples}(e)).
There are no multiples with $Z$ functions because the convolution of spin functions gives zero, as the considered summands have $4\alpha$ and $1\beta$ parts of $\Theta_{1,2,3,4,5}^{3/2 \dagger}$ spin function, and $3\alpha$ and $2\beta$ enter $\Theta_{1,2,3,4,6}^{1/2}$ spin function.
The spin convolution of the latter part is equal to
\begin{gather}
    \label{spin_part_for_Z}
    \displaystyle \Theta_{1,2,3,4,5}^{3/2 \dagger} \beta^{\dagger}_6 \Theta_{1,2,3,4,6}^{1/2} \alpha_5 = \nonumber \\
    = \frac{1}{\sqrt{5}} \left( \alpha_1^{\dagger} \alpha_2^{\dagger} \alpha_3^{\dagger} \alpha_4^{\dagger} \beta_5^{\dagger} + \alpha_1^{\dagger} \alpha_2^{\dagger} \alpha_3^{\dagger} \beta_4^{\dagger} \alpha_5^{\dagger} + \dots \right) \beta_6^{\dagger} \times \nonumber \\
    \times
    \frac{1}{\sqrt{10}} \left( \alpha_1 \alpha_2 \alpha_3 \beta_4 \beta_6 + \alpha_1 \alpha_2 \beta_3 \alpha_4 \beta_6 + \dots \right) \alpha_5 = \nonumber \\
    = \frac{1}{\sqrt{50}} \left( \alpha_1^{\dagger} \alpha_2^{\dagger} \alpha_3^{\dagger} \beta_4^{\dagger} + \alpha_1^{\dagger} \alpha_2^{\dagger} \beta_3^{\dagger} \alpha_4^{\dagger} + \dots \right) \times \nonumber \\
    \times
    \left( \alpha_1 \alpha_2 \alpha_3 \beta_4 + \alpha_1 \alpha_2 \beta_3 \alpha_4 + \dots \right) = \frac{4}{\sqrt{50}} . 
\end{gather}
Then finally we get
\begin{eqnarray}
\label{ae_part_Z_final}
    \displaystyle \mathcal{Z} = |f(0)|^2 \frac{10}{6} \frac{4}{\sqrt{50}} \frac{1}{\sqrt{3}} \frac{4!}{5!} \iint {\rm d}{\bm r_5} {\rm d}{\bm r_6} ~\! \chi^*_6  ~\! \chi_5 \times \nonumber \\
    \times
    \sum\limits_{j=1}^{5} \varphi^{j*}_{5}  \varphi^{j}_{6} ~\! U(|{\bm r_5} - {\bm r_6}|) = \frac{2\sqrt{2}}{\sqrt{3}} \ae |f(0)|^2. \quad
\end{eqnarray}

The calculation of another off-diagonal matrix element $\mathcal{V}$ requires usage of the following ket wave function according to Eq.~(\ref{V})
\begin{gather}
\label{PsiS-1/2_PsiJ1/2}
    \Psi^S_{-1/2} \Psi^J_{1/2} = \nonumber \\
    = \frac{f(0)}{\sqrt{6}} \left\{ \Phi^d_{1,2,3,4,5} \Theta^{-1/2}_{1,2,3,4,5} ~\! \left( \sqrt{\frac{2}{3}} Z_6 \alpha_6 + \sqrt{\frac{1}{3}} \chi^{+}_6 \beta_6 \right) \right. - \nonumber \\
    \left. - (5) - (4) - \dots \right\}.
\end{gather}
As the spin function $\Theta^{1/2}_{1,2,3,4,5}$ contains $3\alpha$ and $2\beta$ in each summation term, while the function $\Theta^{-1/2}_{1,2,3,4,5}$, on the contrary, contains $2\alpha$ and $3\beta$ in each summand, one can see that the exchange matrix element of interaction between quantum states given by Eq.~(\ref{PsiS1/2_PsiJ-1/2}) and Eq.~(\ref{PsiS-1/2_PsiJ1/2}) will reduce to the following expression, having non-zero contributions from only $Z$-orbital terms.
\begin{gather}
\label{ae_part_V}
    \displaystyle \mathcal{V} = |f(0)|^2 \frac{10}{6} \int {\rm d}{\bm r_1} \dots {\rm d}{\bm r_6} ~\! \Phi^{d*}_{1,2,3,4,5} ~\! \Theta_{1,2,3,4,5}^{1/2 \dagger}  \times \nonumber \\
    \times
    \sqrt{\frac{2}{3}} Z^{*}_6 ~\! \beta^{\dagger}_6 ~\! \hat{U} ~\! \Phi^{d}_{1,2,3,4,6} ~\! \Theta_{1,2,3,4,6}^{-1/2} \sqrt{\frac{2}{3}} Z_5 \alpha_5 = \nonumber \\
     =
    |f(0)|^2 \frac{10}{6} \frac{2}{3} \frac{6}{10} \frac{4!}{5!} \iint {\rm d}{\bm r_5} {\rm d}{\bm r_6} ~\! Z^*_6  ~\! Z_5 \times \nonumber \\
    \times
    \sum\limits_{j=1}^{5} \varphi^{j*}_{5}  \varphi^{j}_{6} ~\! U(|{\bm r_5} - {\bm r_6}|) = 2 \ae |f(0)|^2, 
\end{gather}
in which the result of spin functions convolution is included
\begin{gather}
    \label{spin_part_for_V}
    \displaystyle \Theta_{1,2,3,4,5}^{1/2 \dagger} \beta^{\dagger}_6 \Theta_{1,2,3,4,6}^{-1/2} \beta_5 = \nonumber \\
    = \frac{1}{10} \left( \alpha_1^{\dagger} \alpha_2^{\dagger} \alpha_3^{\dagger} \beta_4^{\dagger} \beta_5^{\dagger} + \alpha_1^{\dagger} \alpha_2^{\dagger} \beta_3^{\dagger} \alpha_4^{\dagger} \beta_5^{\dagger} + \dots \right) \beta_6^{\dagger} \times \nonumber \\
    \times
    \left( \beta_1 \beta_2 \beta_3 \alpha_4 \alpha_6 + \beta_1 \beta_2 \alpha_3 \beta_4 \alpha_6 + \dots \right) \alpha_5 = \nonumber \\
    = \frac{1}{10} \left( \alpha_1^{\dagger} \alpha_2^{\dagger} \beta_3^{\dagger} \beta_4^{\dagger} + \dots \right) \left( \alpha_1 \alpha_2 \beta_3 \beta_4 + \dots \right) = \frac{6}{10} . \qquad
\end{gather}


\bibliography{main_paper_prb}

\end{document}